\documentclass[12pt,preprint]{aastex}

\shorttitle{}
\shortauthors{}

\begin{document}

\title{Highly ionized iron absorption lines from outflowing gas 
in the X-ray spectrum of NGC 1365}
\author{
G. Risaliti\altaffilmark{1,2}, S. Bianchi\altaffilmark{3},
G. Matt\altaffilmark{4}, A. Baldi\altaffilmark{1}, 
M. Elvis\altaffilmark{1},  G. Fabbiano\altaffilmark{1},
A. Zezas\altaffilmark{1}
} 
\email{grisaliti@cfa.harvard.edu}

\altaffiltext{1}{Harvard-Smithsonian Center for Astrophysics, 60 Garden St. 
Cambridge, MA 02138 USA}
\altaffiltext{2}{INAF - Osservatorio di Arcetri, L.go E. Fermi 5,
Firenze, Italy}
\altaffiltext{3}{XMM-Newton Science Operations Center, European Space
Astronomy Center, ESA, Apartado 50727, E-28080 Madrid, Spain}
\altaffiltext{4}{Dipartimento di Fisica, Universit\`a degli Studi ``Roma Tre'', 
Via della Vasca Navale 84, I-00146 Roma, Italy}
\begin{abstract}
We present the discovery of four absorption lines
in the X-ray spectrum of the Seyfert Galaxy NGC~1365, at energies
between 6.7 and 8.3~keV. The lines are detected with high statistical
confidence (from $>20\sigma$ for the strongest to $\sim4\sigma$ for
the weakest) in two {\em XMM-Newton} observations 60~ksec long. 
We also detect the same lines, with lower signal-to-noise (but still $>2\sigma$
for each line) in two previous shorter ($\sim10$~ksec) XMM observations.
The spectral analysis identifies these features as FeXXV
and FeXXVI $K\alpha$ and $K\beta$ lines, outflowing with velocities
varying between $\sim1000$ to $\sim5000$~km~s$^{-1}$ among the observations. 
These are the
highest quality detections of such lines  so far. 
The high equivalent widths (EW$(K\alpha)\sim100$~eV) and the
$K\alpha/K\beta$ ratios imply that the lines
are due to absorption of the AGN continuum by a
highly ionized gas with column density $N_H\sim5\times10^{23}$~cm$^{-2}$
at a distance of $\sim50$--$100~R_S$ from the continuum source.
\end{abstract}


\section{Introduction}

Warm absorbers are common among Active Galactic Nuclei (AGNs).
About half of the X-ray spectra of local bright Seyfert galaxies show evidence
of a warm absorber with column densities in the range 10$^{22}$--$10^{23}$~cm$^{-2}$
(e.g., Reynolds 1997).
High resolution grating observations performed with {\em Chandra} and {\em XMM-Newton} 
have shown that the warm absorber has a typical temperature of
$\sim10^6$~K, and, in some cases, outflowing velocities of a few  
$10^3$~km~s$^{-1}$ (Sako et al.~2001, Kaspi et al.~2002, Krongold et al.~2003).
Recently more extreme, ``hot'' absorbers, 
have been discovered through the observation of He-like and H-like iron absorption
lines in several AGNs (e.g., in NGC~3783, Kaspi et al.~2002, Reeves et al.~2004, 
in PG~1115+080, Chartas et al.~2003, in PG~1211+143, Pounds et al.~2003,
in MCG-6-30-15, Young et al.~2005). 

Here we report the discovery of four strong absorption lines in each of four
{\em XMM-Newton} observations 
of the Seyfert Galaxy NGC~1365, in the 6.7--8.3~keV band, which we identify as
FeXXV and FeXXVI $K\alpha$ and $K\beta$ lines.
NGC~1365 (z=0.0055, de~Vaucouleurs et al.~1992) is an X-ray absorbed Seyfert Galaxy, with 
a 2--10~keV flux of the order of 0.5--2$\times10^{-11}$~erg~s$^{-1}$~cm$^{-2}$ 
(Risaliti et al.~2000, 2005, hereafter R05), which shows extreme X-ray variability:
changes from reflection-dominated to transmission dominated states are observed 
on time scale as short as 3~weeks (R05). 
When in a transmission dominated state, the measured cold column
density varies between $\sim1$ and $\sim5\times10^{23}$~cm$^{-2}$.

\section{Data Analysis and Results}

Two {\em XMM-Newton} (Jansen et al.~2001) 
observations of 60~ksec each were obtained in December 2003 and July 2004. 
Two shorter ($\sim10$~ksec) observations were obtained earlier in 2003.
A spectral analysis of all these  observations showed the source to
be in transmission dominated state (R05). 
The observation log is reported in Table~1, together with the main parameters
of the continuum fit which will be described in the next Section.

We reduced the EPIC PN and MOS data using
the SAS 6.0 package. We extracted a spectrum from a circular region of
30" radius, and the background from nearby regions free from bright
serendipitous sources. In all cases the background counts are
negligible at all energies (on average $\sim2$\% of the signal). 
In particular, we checked that no narrow background line\footnote{e.g.,
from Ni, Cu, Zn, see 
http://xmm.vilspa.esa.es/external/xmm\_user\_support/documentation/uhb/node35.html
(Lumb et al.~2002).}
is present in the
energy range which is most crucial for our analysis, i.e., between 6.7 and 8.3~keV,
where the absorption lines are detected.
The data obtained with the two MOS were merged.
Calibration matrices were computed for each spectrum.
We rebinned the spectra in order to have at least 20 counts per bin.
This allows us to use $\chi^2$ minimization in the model fitting.

For all four spectra a good fit 
is obtained with a model consisting of an absorbed power law with $N_H$ between
1 and $5\times10^{23}$~cm$^{-2}$,
plus a cold reflection component (using the PEXRAV model in XSPEC,
Magdziarz \& Zdziarski 1995) and a narrow emission line at
6.4~keV. 
At energies below the photoelectric
cut-off a thermal emission component dominates, with $kT\sim0.8$~keV (R05).
A broad emission line,
probably relativistic, is also needed to properly fit the 4--6~keV
region\footnote{Alternative continuum models are possible, for
example with the inclusion of partial
covering.
Here we do not investigate this point since we are only interested in
an analytic fit to the continuum, in order to have a correct determination
of the narrow absorption lines parameters.}. In addition to these
``standard'' components, four absorption lines in the energy range
6.7--8.3~keV are strongly requested by the fit.
A complete analysis of the emission spectrum of NGC~1365
will be presented elsewhere, together with a detailed study of the
spectral variability during the two long observations.
Here we concentrate on the most striking 
features in these spectra, i.e., the group of absorption lines present in
the 6.7--8.3~keV spectral region. 
In order to visually show the relevance of these lines,
in Fig.~1 we plot the 5--10~keV residuals to 
a model with the same components as above, except for the four absorption lines, 
and fitted ignoring the energy ranges 6.7--7.2~keV
and 7.8--8.3~keV. The spacing of the lines strongly suggests that they
are FeXXV and FeXXVI $K\alpha$ and $K\beta$.

Accordingly, in the global 0.5--10~keV fits the four lines were fitted with narrow
Gaussians, with free energies 
allowed for each single line and each single fit.
All the four lines are significantly detected in all spectra, except in 
two cases where the detection is marginal (FeXXV~K$\beta$ in OBS~1 and OBS~2).
For the two highest quality observations, OBS~3 and OBS~4, the fits were
performed on PN and MOS data separately. 

Since the best fit energies are
close to those of the four lines mentioned above, we estimated the blueshift/redshift
velocities corresponding to the differences between the measured and
theoretical energies for these lines.
The velocities are shown in Fig.~2. In both
observations the results are in agreement with a common outflow velocity for the four lines.
Therefore, we performed a
new fit fixing the energies of the four lines to the theoretical values (taking
into account for the redshift of the source), and
allowing for a single free redshift/blueshift.
We obtained as good a fit as in the previous case where
all the lines were left free ($\Delta(\chi^2)<8$ in the global fit in all observations, 
with three less free parameters). 

The best fit line parameters are listed in Table~2. The continuum parameters,
listed in Table.~1, are typical of X-ray spectra of obscured, Compton-thin 
Seyfert galaxies, and will be discussed in detail in a forthcoming paper.

In Fig.~3 we show the 5-10~keV spectrum and best fit model for the two
long observations.
The measured $K\alpha/K\beta$ ratios (Table~3)
are significantly smaller than those expected from the oscillator strengths 
ratios\footnote{
Throughout this paper we refer to the atomic physics values of
the NIST Atomic database (http://physics.nist.gov), and reported
in Table~1 of Bianchi et al. 2005,
hereafter B05.}.


In all four spectra the lines are blueshifted with respect to the expected 
theoretical line energy peaks. The velocity shifts are between 1,000 and 5,200~km~s$^{-1}$ 
(Table~2), and are statistically highly significant, except for the first observation,
where the data are in agreement with zero velocity. 
The differences between the outflow velocities
are highly significant, and cannot be due to instrumental effects. 
In order to check the reality of this shift, we compared
the low energy thermal emission lines in the $\sim1$~keV region (which 
are expected to be constant, given the good S/N and the extent of the thermal emission region resolved 
by {\em Chandra}, R05) and found that they overlap perfectly ($\Delta(E)<10$~eV).



\section{Discussion}

The main results of our spectral analysis are the following:\\
$\bullet$ We detected four absorption lines in the energy range
6.7--8.3~keV in four {\em XMM-Newton} observations of NGC~1365,
with equivalent width in the 50-150~eV range.
We identified these lines as FeXXV and FeXXVI $K\alpha$ and $K\beta$.\\
$\bullet$ In all observations the lines are blueshifted. Blueshift
velocities are the same for all lines in the same observation, but
vary among the different observations 
from $\sim1,000$ to $\sim5,000$~km~s$^{-1}$.\\
$\bullet$ The $K\alpha/K\beta$ ratios are much smaller than
the theoretical oscillator strength values (Table~3), and are
consistent with staying constant between the observations.


In the following we discuss the composition (ionization state, column
density), the geometrical structure, and origin  
of the hot absorber. 

{\bf Composition.} 
The low $K\alpha/K\beta$ ratios can
only be explained if saturation effects are significant. Any other
physical explanation is ruled out, since (a) the theoretical ratio
is well established, since it involves transitions to the
fundamental level in relatively simple (He-like and H-like) atomic species,
and (b) the energy difference between the lines is too small
for the continuum slope to play a significant role: for a photon
index $\Gamma=2.0$ we have $F_E$(6.7~keV)$/F_E$(7.9~keV)$=1.17$.
Following B05, an equivalent width of the
$K\alpha$ lines EW$\sim$100--150~eV
can be obtained with $N_H>10^{23}$~cm$^{-2}$, 
a turbulent velocity of the absorbing gas $v_t=1,000$~km~s$^{-1}$,
and an ionization parameter
$\log U_X=0$.\footnote{We adopt the definition in B05: 
$U_X=(\int^{10}_2\frac{L_\nu}{h\nu}d\nu)/(4\pi r^2cn_e)$, similar to
the one first introduced by Netzer~(1996). For ease 
of comparison, if $\log U_X=0$, 
the ionization parameter defined as $U_X'=L_X/(n_er^2)$ has
a value $U_X'\sim5,000$ (for an ionizing continuum with $\Gamma=2$).
}
We note that the EPIC resolution at 8~keV corresponds
to velocities of $\sim5,000$~km~s$^{-1}$. With the available statistics
we find that the minimum detectable velocity is of the order of 2,000~km~$^{-1}$.
Therefore, no useful constraint on $v_t$ can be obtained from our fits.

We calculated the 
curve of growth for the $K\beta$ lines using the same model as
B05.
In order to have a precise estimate of the EW ratios, we also
considered the contribution of the FeXXV $K\gamma$ line at 8.29 keV, 
which is blended with the FeXXVI $K\beta$ in low resolution spectra, and has an
oscillator strength about half of that of the FeXXVI $K\beta$.
A further minor contribution could come from the
NiXXVII (He like) $K\alpha$ at 7.98~keV. Its oscillator strength
is 0.19 that of FeXXV $K\beta$. An overabundance of nickel could
make this contribution important. However, we note that in this case
the line profile would be significantly altered, given that the energy
separation between the two lines ($\Delta E=80$~eV) is comparable
to the EPIC resolution ($\sim150$~eV for both PN and MOS cameras; but note that 
with the high statistics available the peak energies of the lines is 
determined with a much better precision, $\sim$10--15~eV). 
This is not observed
(Fig.~1). We conclude that the contribution of nickel lines cannot be
significant.
In Fig.~4 we plot the curve of growth for the FeXXV and FeXXVI $K\alpha$ lines
(from B05) and the $K\alpha/K\beta$ ratio.  
Note that both the large EWs and the low $K\alpha/K\beta$ ratios require
an absorber with 
a column density $N_H\sim5\times10^{23}$~cm$^{-2}$ and a turbulent
velocity  $\sigma\sim$500--1,000~km~s$^{-1}$.
The right panels of Fig.~4 show the importance of saturation effects
in the $K\alpha/K\beta$ ratios.
High column densities are required in order to reproduce the
small observed ratios. 

The above estimates are computed adopting a fixed ionization 
parameter $U_X$. Therefore we expect they slightly change for 
different choices of $U_X$. However we do not expect 
to find acceptable solutions for significantly lower or higher 
ionization parameters: if (a) $\log U_X>>0$  
the iron atoms are almost completely stripped, making impossible to obtain
as high EWs as observed; if (b) $-1 < \log U_X < 0$ the He-like Fe atoms
are overabundant with respect to the H-like ones, because the ionizing
continuum is enough to strip all the L shell electrons, but not to 
strip the 2 inner electrons. The ratio between FeXXV and FeXXVI lines
would then be much higher than observed. This argument is discussed 
in more detail in B05.

The turbulent velocity is significantly higher than the thermal velocity
of a gas with the same ionization state ($\sigma_T\sim$100--200~km~s$^{-1}$). 
As a consequence, the line widths probably indicate true bulk
motions. These might be related to the observed changes in outflow 
velocity.

{\bf Geometrical properties}. The high ionization parameter required
to explain the line ratios can be used to put constraints on
the distance of the line absorber from the X-ray continuum source,
assuming that the ionization state is due to UV radiation
from the central source, rather than to a high thermal temperature of the gas. 
For $\log U_X=0$, $N_H=5\times10^{23}$~cm$^{-2}$, and a 2--10 keV
photon index $\Gamma=2$ we obtain $R\sim2\times10^{15}~(\Delta R/R)~L_{42}$~cm, 
where $L_{42}$ is the 2--10~keV luminosity in units of 10$^{42}$~erg~s$^{-1}$ (Table~1)
and $\Delta R$ is the thickness of the absorber along the radial direction.
$R$ is the average distance of the absorber, so we expect $\Delta R/R \sim 1$.
We note that a distance of $\sim10^{15}$~cm is of the order of 
what a gas with a velocity 
$v\sim4,000$~km~s$^{-1}$ (intermediate between the two measurements in
the last two observations, OBS~3 and OBS~4) 
would cover in $\sim6$~months between OBS~3 and OBS~4 ($\sim2\times10^7$~sec). 
This is in qualitative agreement with our assumption of an absorbing gas
with a thickness of the same order of the distance from the central source.
A smaller thickness ($\Delta R/R <1$) would imply an even smaller distance
from the central source.

{\bf Origin}. The short distance inferred from the high ionization state of the
absorber, and the rough black hole mass estimate ($M_{BH}\sim10^8~M_\odot$, R05)
imply that the gas is located at the radii of the accretion disk,
$\sim50~(M/M_8)^{-1}$ Schwarzschild radii from the black hole. Hot gas could be present
in this region either as the external part of the hot corona believed to be responsible of
the X-ray emission in AGNs\footnote{The inner
corona, emitting the bulk of X-ray radiation, has to have $kT\sim100$~eV,
therefore it is too hot to absorb its own radiation.}
 (e.g., Haardt \& Maraschi 1991), or as the inner part
of a wind arising from the disk (Murray \& Chiang~1995, Proga et al.~2000). 

We note that the observed blueshift velocity is smaller than the escape velocity
at 50$~(M/M_8)^{-1}R_S$, which is of the order of about 50,000$~(M/M_8)^{-0.5}$~km~s$^{-1}$. 
However, the observed velocity is of the order of that predicted 
in the vertical part of the funnel-shaped wind of Elvis (2000).
The main problem  with the wind 
scenario is that radiation force in not an effective acceleration
mechanism in our case, since the gas is overionised (so line absorption is not
an efficient way to transfer momentum from radiation to gas) and the bolometric
luminosity (of the order of a few $10^{43}$~erg~s$^{-1}$ adopting a standard X-ray to bolometric
correction, Risaliti \& Elvis~2004) is a small fraction of the Eddington luminosity,
L$_{EDD}\sim2\times10^{46}(M/M_8)^{-1}$~erg~s$^{-1}$, 
making Thomson scattering acceleration negligible with respect
to gravity. However, two mechanisms could provide the observed highly ionized gas:
(a) a poloidal magnetic field could effectively extract the gas from the 
accretion disk and form a wind (Blandford \& Payne 1982, Konigl \& Kartje 1994);
(b) gas arising from the accretion disk due to local instabilities would be ionized
by the central UV/X-ray source, and initially expand towards the vertical direction
(with respect to the disk plane) due to its internal pressure. The gas would then 
fail to form a wind if no external force is present, but a fraction of
the line of sight, at small angles with respect to the disk plane, would be covered by
such gas, with the required ionization state and column density, and with a
(temporary) outflowing velocity of the order of a few $10^3$~km~s$^{-1}$. Similar properties
are predicted for the inner ``hitchhiking gas'' in radiation driven wind models
(Murray \& Chiang 1995). 

\acknowledgements
We are grateful to F. Nicastro and N. Brickhouse for
useful discussions.
This work was partially supported by NASA grants NAG5-13161,
NNG04GF97G, and NAG5-16932.


\clearpage

\begin{figure}
\epsscale{0.8}
\plotone{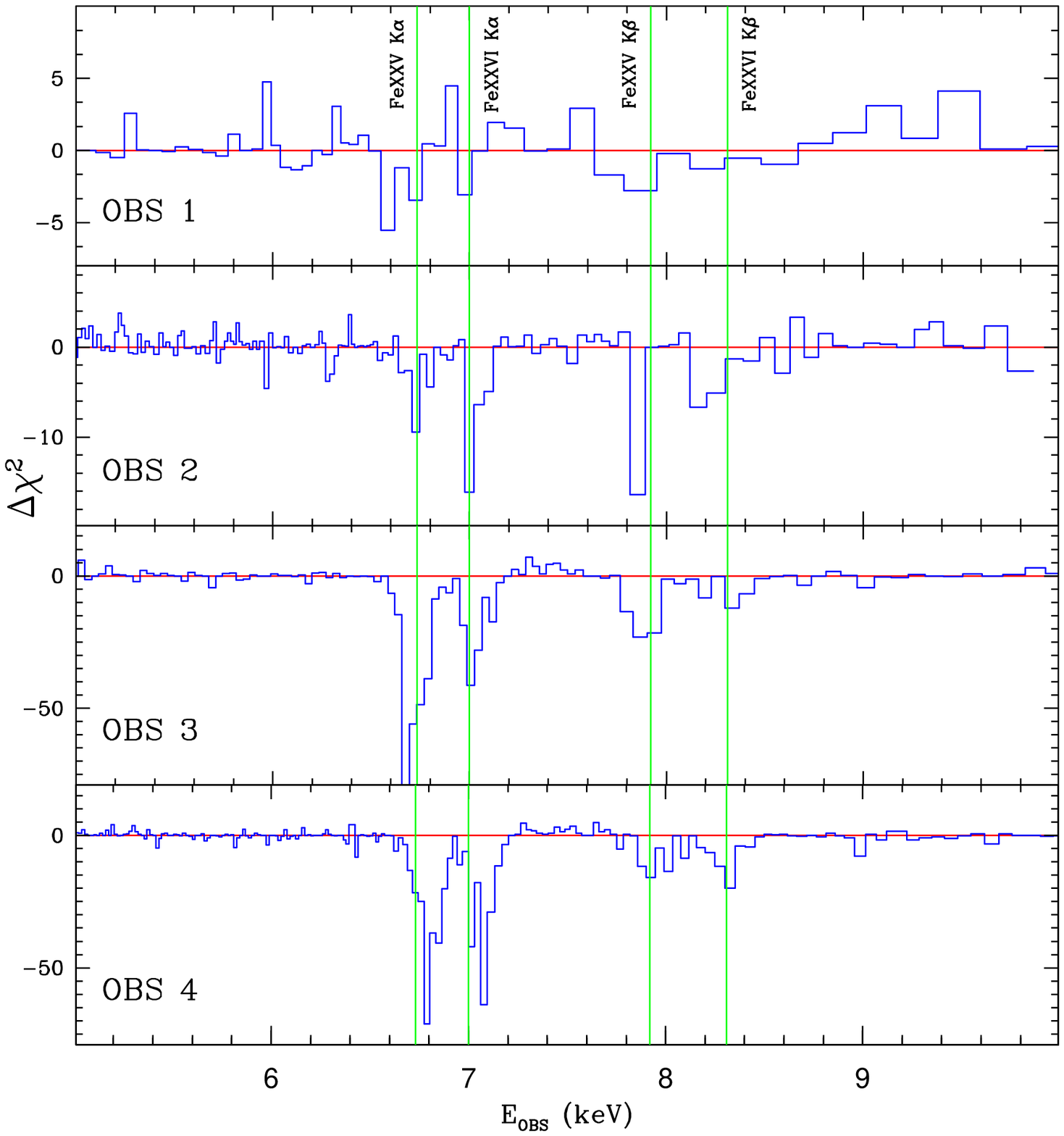}
\caption{Contributions to $\chi^2$ in the 5-10 keV range for the four
MOS observations of NGC~1365. The continuum model has been
fitted ignoring the energy intervals 6.7-7.2~keV and 7.8-8.3~keV.
The four vertical lines show the rest frame energy of the four lines
FeXXV~K$\alpha$, FeXXVI~K$\alpha$, FeXXV~K$\beta$, FeXXVI~K$\beta$.}
\end{figure}

\clearpage

\begin{figure}
\epsscale{0.8}
\plotone{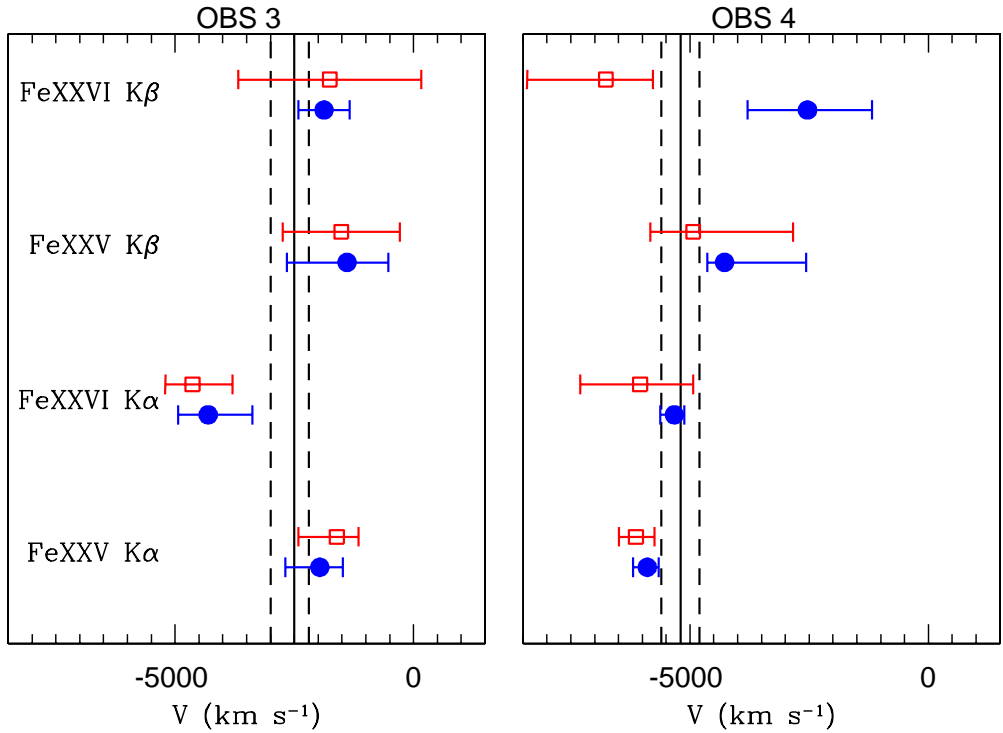}
\caption{
Best fit velocities for the four iron absorption lines in the
two highest quality observations, OBS~3 and OBS~4. For each line we separately
plot the results from the PN spectra (blue circles) and the
MOS spectra (red squares). The vertical solid line shows the best fit
value obtained requiring a common velocity for the four lines with
$\pm1~\sigma$ values shown as vertical dashed lines. 
}
\end{figure}

\clearpage

\begin{figure}
\epsscale{0.8}
\plotone{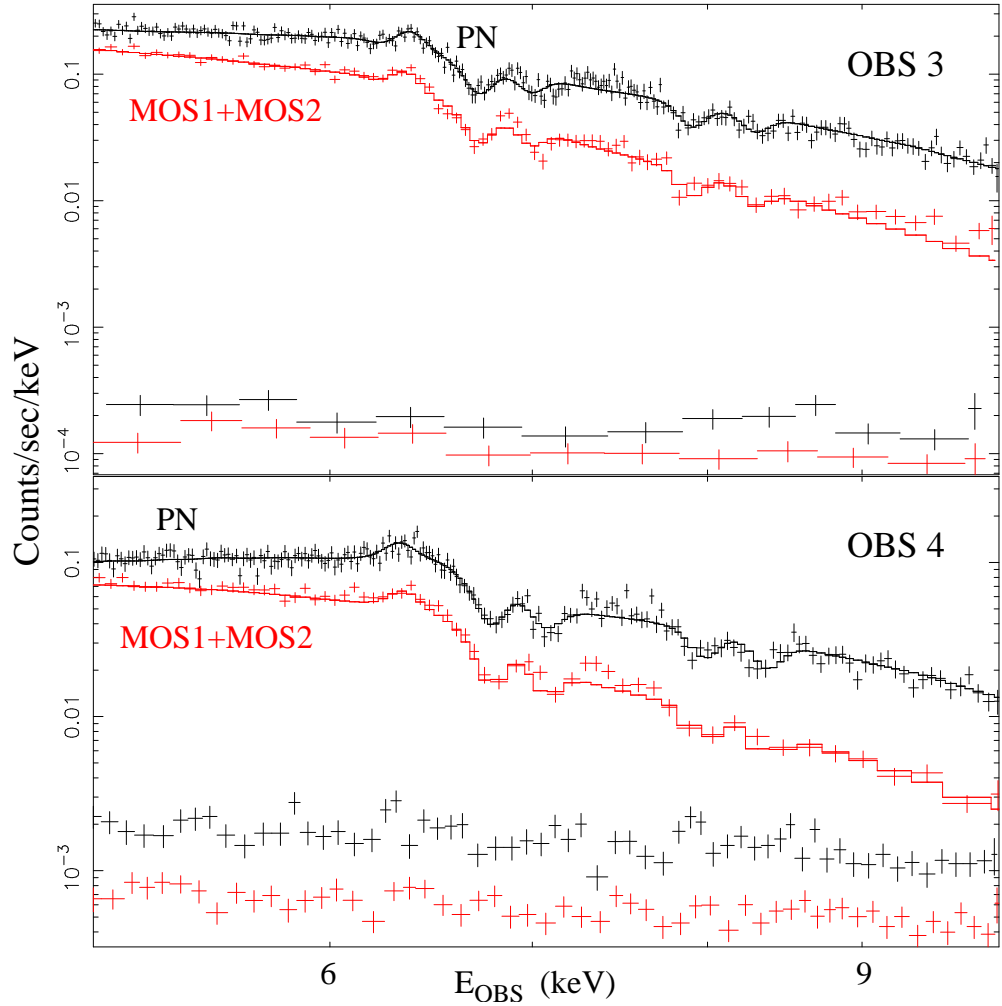}
\caption{Hard X-ray (5-10~keV) spectra and best fit model for the two
long {\em XMM-Newton} observations of NGC~1365. The background spectrum
is shown for each observation.
}
\end{figure}

\clearpage

\begin{figure}
\epsscale{0.9}
\plotone{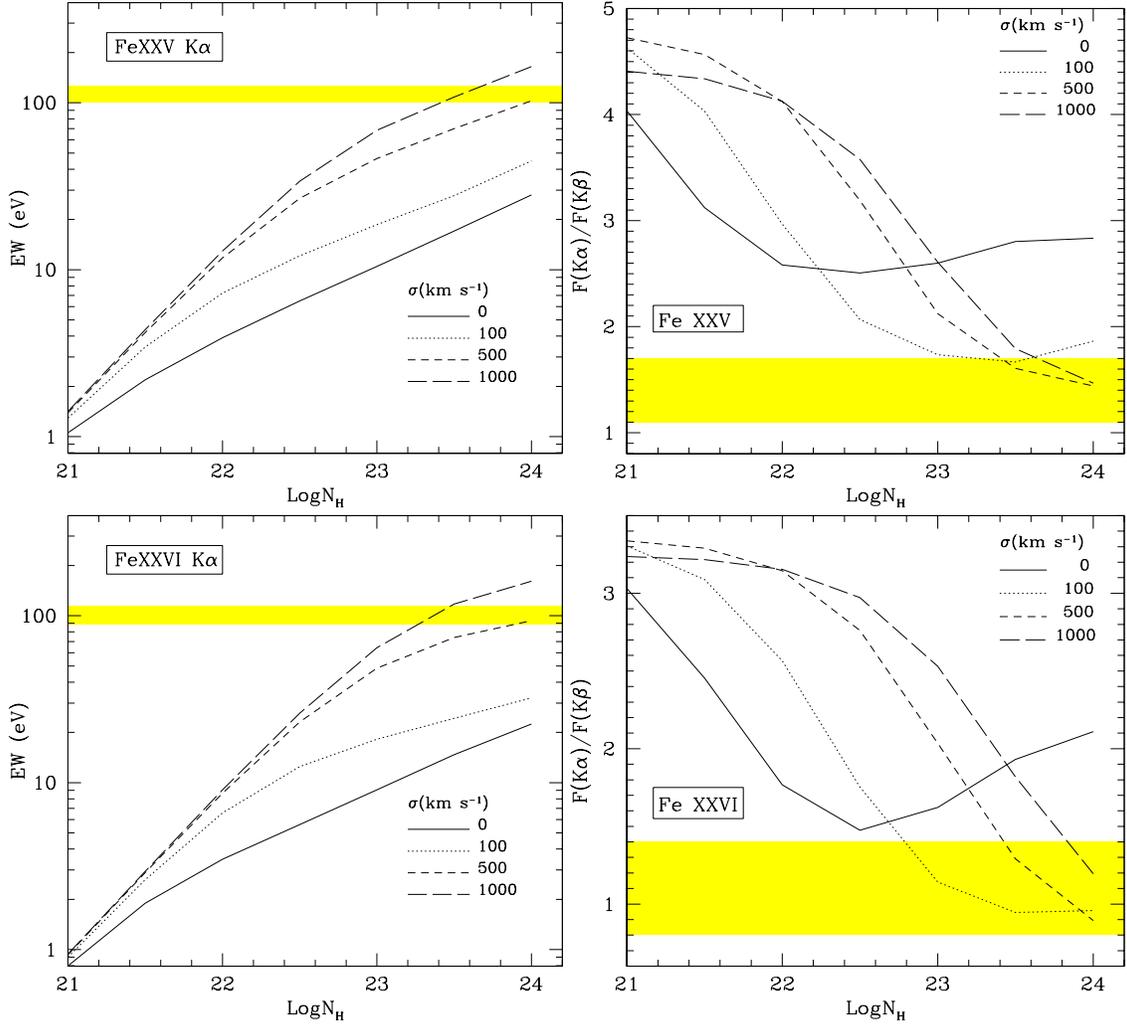}
\caption{Left panels: curves of growth for the FeXXV and FeXXVI 
$K\alpha$ lines, for $\log U_X=0$. Right panels: $K\alpha/K\beta$ ratios for FeXXV and
FeXXVI. The yellow shaded regions are the 90\% confidence intervals
for the averages obtained from the four observations.
}
\end{figure}

\clearpage

\begin{table}
\caption{Observation Log and Spectral Parameters}
\centerline{\begin{tabular}{cccccccc}
\tableline
\tableline
OBS      & Date     & t\tablenotemark{a} & Counts\tablenotemark{b}& F\tablenotemark{c} &L\tablenotemark{d} &  $\Gamma$ & $N_H$\tablenotemark{e}\\
\tableline
 1 & 2003 Jan 16 & 14.8  & 14866     & 0.48 & 1.0 & 2.15$^{+0.03}_{-0.06}$ & 51$^{+3}_{-3}$\\
 2 & 2003 Aug 13 & 7.1   & 9466      & 0.73 & 1.4 & 2.32$^{+0.12}_{-0.02}$ & 37$^{+2}_{-2}$  \\
 3 & 2004 Jan 17 & 59    & 121967    & 1.37 & 1.7 & 2.55$^{+0.05}_{-0.03}$ & 16.3$^{+0.2}_{-0.2}$ \\
 4 & 2004 Jul 24 & 60    & 74056     & 0.76 & 0.8 & 2.11$^{+0.03}_{-0.04}$ & 29.4$^{+0.3}_{-0.3}$\\
\tableline
\end{tabular}}
\footnotesize{
\tablenotetext{a}{Net observing time in ksec.} 
\tablenotetext{b}{Source counts in the 0.5-10~keV interval.}
\tablenotetext{c}{2-10~keV observed flux, in units of $10^{-11}$~erg~s$^{-1}$~cm$^{-1}$.}
\tablenotetext{d}{Intrinsic 2-10~keV luminosity in units
of $10^{42}$~erg~s$^{-1}$.}
\tablenotetext{e}{In units of $10^{22}$~cm$^{-2}$.}}
\end{table}

\clearpage

\begin{table*}
\caption{NGC 1365 - Data on absorption lines}
\centerline{\begin{tabular}{lccccccccc}
\tableline
\tableline
Line            & E(keV)     & \multicolumn{2}{c}{OBS 1} & 
		               \multicolumn{2}{c}{OBS 2} & 
                               \multicolumn{2}{c}{OBS 3} & 
                               \multicolumn{2}{c}{OBS 4} \\
                &        & EW (eV)\tablenotemark{a} & $\Delta(\chi^2)$\tablenotemark{c} & EW (eV)\tablenotemark{a} & $\Delta(\chi^2)$\tablenotemark{c} & 
                           EW (eV)\tablenotemark{a} & $\Delta(\chi^2)$\tablenotemark{c} & EW (eV)\tablenotemark{a} & $\Delta(\chi^2)$\tablenotemark{c} \\
\tableline
FeXXV $K\alpha$ & 6.697  & 110$^{+45}_{-70}$ & 17    & 135$\pm30$         &30    &
                           120$\pm20$        & 173   & 154$\pm20$         &264   \\
FeXXVI $K\alpha$& 6.966  & 94$^{+50}_{-50}$  & 11    & 111$\pm$30         &20    &
                           104$^{+15}_{-20}$ & 52    & 145$^{+15}_{-20}$  &167   \\
FeXXV $K\beta$  & 7.880  & 81$\pm55$         & 5     & 100$\pm60$         &6     &
                           93$\pm10$         & 43    & 108$\pm20$          &60    \\
FeXXVI $K\beta$\tablenotemark{d} & 8.268  & 127$\pm60$        & 12    & 130$\pm60$         &10    &
                           80$\pm15$         & 18    & 110$\pm25$          &59    \\
V(km~s$^{-1}$)\tablenotemark{b} && \multicolumn{2}{c}{1000$^{+1000}_{-1500}$}          &
		     \multicolumn{2}{c}{2000$^{+1800}_{-400}$} &
		     \multicolumn{2}{c}{2500$^{+500}_{-300}$}  &
		     \multicolumn{2}{c}{5200$^{+400}_{-400}$}  \\
\tableline
\end{tabular}}
\footnotesize{
\tablenotetext{a}{Errors are at a 90\% confidence level.}
\tablenotetext{b}{Outflowing velocity of the absorbing gas.}
\tablenotetext{c}{$\Delta(\chi^2)$ with
respect to the best fit model without the four absorption lines. The four lines 
are forced to have the same blueshift, so the model including the lines
has 5 more parameters than the no-line model.} 
\tablenotetext{d}{The FeXXVI $K\beta$ line is blended
with the FeXXV $K\gamma$ (see text for details).} 
}
\end{table*}

\clearpage

\begin{table}
\caption{FeXXV and FeXXVI $K\alpha/K\beta$ ratios}
\centerline{\begin{tabular}{lcccccc}
\tableline
\tableline
       & OBS 1 & OBS 2 & OBS 3 & OBS 4 & R$_{AV}^a$ & Th.$^b$ \\
\tableline
FeXXV & 1.4$\pm$1.7 & 1.4$\pm$1.1 & 1.4$\pm$0.4 & 1.4$\pm$0.4 & 1.4$\pm$0.3 & 5.0\\
FeXXVI& 0.7$\pm$0.7 & 0.9$\pm$0.6 & 1.4$\pm$0.7 & 1.3$\pm$0.4 & 1.1$\pm$0.3 & 3.2\\
\tableline
\end{tabular}}
\footnotesize{
$^a$: Average ratios: 
$^b$: Oscillator strengths ratio. In R2 the contribution of FeXXV K$\gamma$
has been taken into account.}
\end{table}

\end{document}